\documentclass[conference]{IEEEtran}
\usepackage{algorithm}
\usepackage[noend]{algpseudocode}
\usepackage{fancyhdr}
\usepackage{epsfig,latexsym}
\usepackage{float}
\usepackage{indentfirst}
\usepackage{amsmath}
\usepackage{amssymb}
\usepackage{times}
\usepackage{subfigure}
\usepackage{pifont}
\usepackage{psfrag}
\usepackage{cite}
\usepackage{url}
\usepackage{lastpage}
\usepackage{bm}
\usepackage{color}
\usepackage{fancyhdr}
\usepackage[center]{caption2}
\usepackage{balance}
\usepackage[left=0.63in, right=0.63in, top=0.76in, bottom=1.01in]{geometry}

\newtheorem{Proposition}{Proposition}

\ifCLASSINFOpdf  
\else
\fi
\begin{document}
\title{Robust Beamforming for AN Aided MISO SWIPT System with Unknown Eavesdroppers and Non-linear EH Model}

\author{\IEEEauthorblockN{Miao Zhang$^{*}$, Kanapathippillai Cumanan$^{*}$, Lei Ni$^{\dagger}$, Hang Hu$^{\ddagger}$, Alister G. Burr$^{*}$ and Zhiguo Ding$^{\S}$.}
	\IEEEauthorblockA{$^{*}$Department of Electronic Engineering, University of York, York, YO10 5DD, United Kingdom\\
		$^{\dagger}$Graduate School, Air Force Engineering University, Xi’an, 710077 China\\$^{\ddagger}$College of Information and Navigation, Air Force Engineering University, Xi’an, 710077 China\\ $^{\S}$The University of Manchester, Manchester, M13 9PL, United Kingdom \\
		Email: $^{*}$\{mz1022, kanapathippillai.cumanan, alister.burr\}@york.ac.uk, 
	$^{\dagger}$nileikgd@163.com, $^{\ddagger}$xd\_huhang@126.com,\\
	$^{\S}$zhiguo.ding@manchester.ac.uk.}}


%Telephone: (800) 555--1212, Fax: (888) 555--1212}


\maketitle

\begin{abstract}
This work studies a beamforming design for downlink transmission of a multi-user multiple-input single-output (MISO) system where each legitimate user employs a power splitting (PS) based simultaneous wireless information and power transfer (SWIPT) technique. The transmitter intends to send confidential information to its legitimate users in the presence of purely unknown eavesdroppers. Since the transmitter does not have any knowledge of the eavesdroppers' channel state information (CSI), we consider an artificial noise (AN) approach to establishing secure communication. This beamforming design is developed by maximizing the AN power to interfere with the eavesdropper as much as possible. Based on the assumption of imperfect CSI of legitimate users at the transmitter, two robust design approaches for the joint beamforming and PS ratio have been studied to maximize the AN power under both energy harvesting (EH) and signal-to-interference-plus-noise ratio (SINR) requirements at each legitimate user. In the first robust design, we consider the bounded channel uncertainties, and employ semidefinite relaxation (SDR) and a linear matrix inequality (LMI) representation to transform the original problem into a semidefinite program (SDP). In the second robust design, we consider the statistical channel uncertainties, and show that the proposed problem can be reformulated into another form of SDP through both SDR and Bernstein-type inequality. In addition, the non-linear energy harvesting (EH) model is incorporated in this work as it could reflect the characteristics of practical radio frequency(RF)-EH conversion circuit. Simulation results have been provided to demonstrate the performance of our proposed robust designs.
\end{abstract}

% \fi
\IEEEpeerreviewmaketitle

\section{Introduction}
Recently, simultaneous wireless information and power transfer (SWIPT) has been recognized as a promising solution to transmit information and energy to incorporated wireless devices. In addition, SWIPT has been considered as a potential technique to extend the battery lifetime of wireless networks, especially in wireless sensor networks and Internet of Things (IoT) \cite{zhang2013mimo,lu2017swipt}. In contrast to the conventional energy harvesting techniques which extract energy from natural energy sources such as solar and wind, SWIPT can be easily implemented on wireless devices \cite{ng2014robust,zhang2016secrecy,ng2015secure} by exploiting the radio frequency (RF) signals \cite{varshney,grover2010shannon}.  In a SWIPT system, the receivers can perform both information decoding (ID) and energy harvesting (EH) simultaneously \cite{lu2017swipt}. However, SWIPT systems are more vulnerable against eavesdropping, due to the fact that the energy receivers can perform both EH and decoding the confidential information intended for legitimate users \cite{chu2017robust}. Recently, information theoretic based physical layer security techniques have received a considerable attention in enhancing security in SWIPT systems \cite{chu2016simultaneous, liu, ng2015secure, zhang2016secrecy}. Unlike conventional cryptographic techniques that completely rely on the computational complexity of some mathematical problems \cite{liang2009information, zhang2017secure,chu2014secrecy}, physical layer security techniques are developed by exploiting physical layer characteristics of wireless channels \cite{Shannon,Wyner,csiszar,cumanan2017physical,cumanan2016secrecy,cumanan2017secure, cumanan2016secure}. In SWIPT systems, an artificial noise (AN) approach can be utilized to degrade the interception capability of energy receivers while increasing the amount of harvested energy. The AN aided beamforming designs for establishing secure transmission in SWIPT systems have been studied in  \cite{chu2017robust, li2014robust, ng2014robust,zhu2016joint}.

Most existing works that adopt the linear EH model assume that the output direct current (DC) power is independent of the input power. However, in practice, the EH circuit results in a non-linear power conversion due to the rectifier in the RF-EH conversion circuit which is the key element in wireless power transfer implementation \cite{valenta2014harvesting}. Therefore, the assumption of a linear EH model in the literature may not be able to incorporate the non-linear characteristics in practical EH scenarios \cite{boshkovska2016power}. Recently, the practical parametric non-linear EH model has been considered in \cite{boshkovska2016power}. In contrast to the linear EH model, the non-linear one includes the characteristics of practical  RF-EH conversion circuits. A power allocation strategy for a SWIPT system with non-linear EH model is proposed in \cite{boshkovska2016power}, whereas EH maximization is investigated with a non-linear EH model for multiple antenna systems in \cite{boshkovska2015practical, boshkovska2016robust, lu2017swipt, niu2017robust,wang2017wirelessly}.

In most existing works on secure transmission schemes, it is assumed that the transmitter has perfect or partial knowledge of the eavesdropper's CSI \cite{lu2017swipt, ng2014robust,chu2017robust,cumanan2008robust, cumanan2016666}. However, it is not always possible to obtain the eavesdropper's CSI in practical scenarios, for example, the eavesdroppers might be purely passive during some transmissions. Most recently, robust beamforming designs with unknown eavesdroppers have been studied in \cite{xiong2016robust, ma2017cooperative}, where the power of AN or jamming signal is maximized to confuse the unknown eavesdroppers as much as possible under the QoS requirement at the legitimate user. 

Motivated by the aforementioned aspects, we consider a MISO secure SWIPT system with power splitting (PS) in this paper, where the transmitter and legitimate users are equipped with multiple and single antennas, respectively. It is assumed that all the legitimate users can simultaneously process information decoding and energy harvesting. A non-linear EH model is adopted in the legitimate users to capture the characteristics of practical EH circuits. In addition, it is assumed that all the eavesdroppers are purely passive, hence their CSI is not available at the transmitter, and we exploit AN to mask the signal intended for the legitimate users. In this work, our aim is to maximize the power of AN to confuse the unknown eavesdroppers while satisfying quality of service and EH requirements at the legitimate users. We develop two robust joint beamforming and PS designs with the AN approach: (1) For the bounded channel uncertainties, we show that the problem can be transformed as a semidefinite program (SDP) through semidefinite relaxation (SDR) and a linear matrix inequality representation; (2) For the statistical channel uncertainties, we exploit both SDR and Bernstein-type inequality to reformulate the original problem into another SDP.

The remainder of this paper is organized as follows. The system model is presented in Section II, whereas the two robust joint beamforming and PS designs are developed in Section III. Section IV provides simulation results to validate the effectiveness of our proposed designs and finally Section V concludes this paper. 
\subsection{Notations}
We use the upper and lower case boldface letters for matrices and vectors, respectively. $(\cdot)^{-1}$, $(\cdot)^T$ and $(\cdot)^H$ stand for inverse, transpose and conjugate transpose operation, respectively. $\mathbf{A}\succeq\mathbf{0}$ means that $\mathbf{A}$ is a positive semidefinite matrix. $||\mathbf{A}||$ represents the Euclidean norm of matrix $\mathbf{A}$. The $\textrm{rank}(\mathbf{A})$ and $\textrm{tr}(\mathbf{A})$ denotes the rank and the trace of matrix $\mathbf{A}$, respectively. The circularly symmetric complex Gaussian (CSCG) distribution is represented by $\mathcal{CN}(\mu,\sigma^2)$ with mean $\mu$ and variance $\sigma^2$. $\mathbb{H}^{N}$ denotes the set of all $N \times N$ Hermitian matrices. $Re(\cdot)$ extracts the real part of a complex number, whereas $\textrm{vec}(\cdot)$ denotes the vector by stacking columns of a matrix.
\section{System Model}
In this work, we consider the downlink transmission of an AN aided MISO SWIPT system with $K$ legitimate users and $J$ eavesdroppers as shown in Fig. 1. It is assumed that the legitimate transmitter is equipped with $N_{T}$ antennas whereas both legitimate users and eavesdroppers are equipped with single antennas. The channels between the transmitter and the $k$-th legitimate user, and the $j$-th eavesdropper are denoted as $\mathbf{h}_{s,k}\in\mathcal{C}^{N_{T}\times1}$ and $\mathbf{h}_{e,j}\in\mathcal{C}^{N_{T}\times1}$, respectively. $\mathbf{q}_{k}\in\mathcal{C}^{N_{T}\times1}$ and $\mathbf{v}\in\mathcal{C}^{N_{T}\times1}$ represent the information signal vector and the artificial noise vector. In addition, the AN vector $\mathbf{v}$ follows zero mean complex Gaussian distribution with covariance matrix $\mathbf{V}\succeq 0$. The received signal at the $k$-th legitimate user and the $j$-th eavesdropper can be expressed as 
\begin{center}
	\begin{figure}[ht!]
		\includegraphics[width=\linewidth]{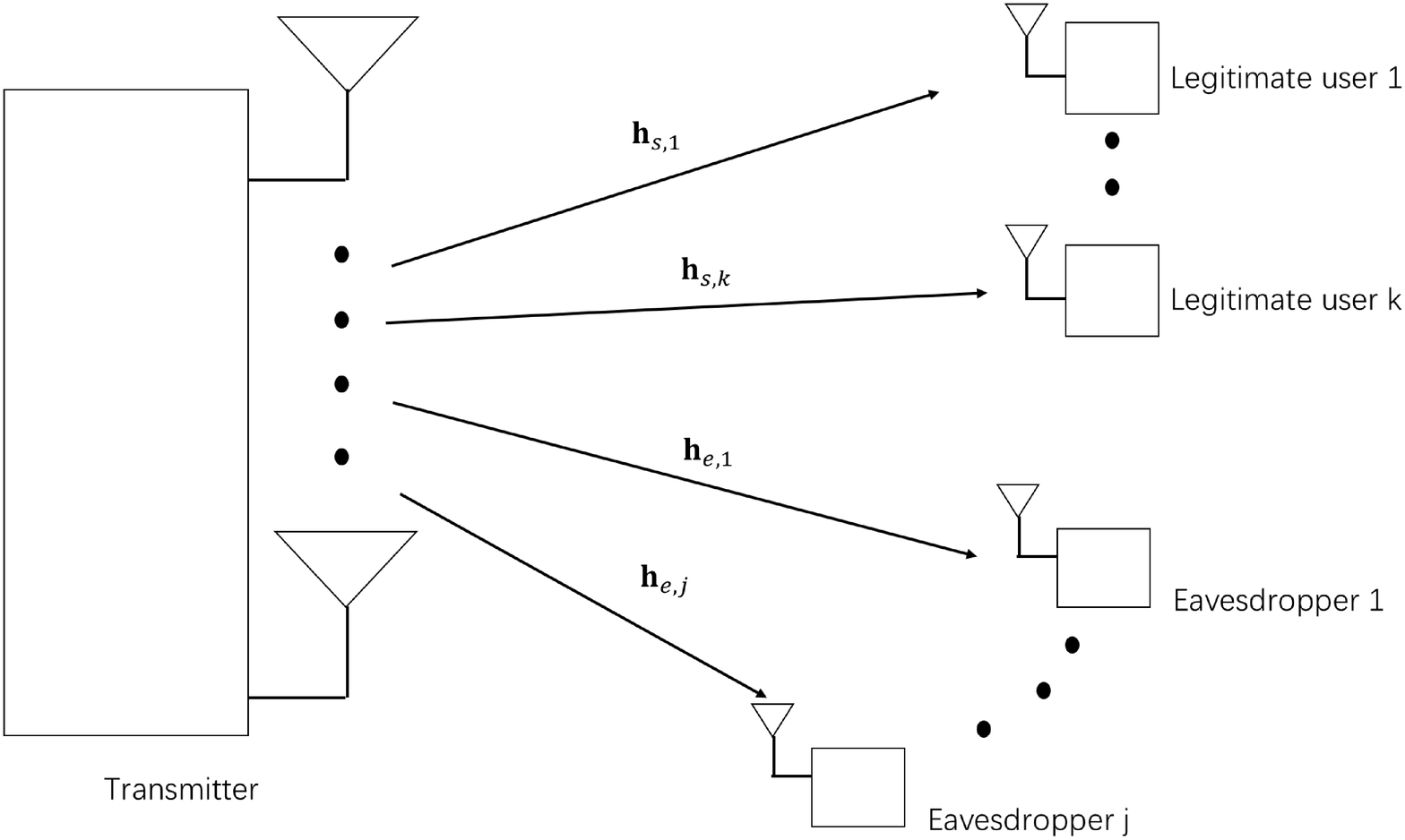}
		\caption{A MISO SWIPT system with $K$ single-antenna legitimate users in the presence of $J$ single-antenna eavesdroppers.}	
		\label{fig:SRP}
	\end{figure}
\end{center}
\begin{align}
y_{s,k}=\sum_{l=1}^{K}\mathbf{h}_{s,k}^{H}\mathbf{q}_{l}+\mathbf{h}_{s,k}^{H}\mathbf{v}+n_{s,k},
\end{align}
\begin{align}
y_{e,j}=\sum_{l=1}^{K}\mathbf{h}_{e,j}^{H}\mathbf{q}_{l}+\mathbf{h}_{e,j}^{H}\mathbf{v}+n_{e,j},
\end{align}
respectively, where $n_{s,k}$ and $n_{e,j}$ are the joint effects of thermal noise and signal processing noise, respectively, at the $k$-th legitimate receiver and the $j$-th eavesdropper, which are modelled as additive white Gaussian noise (AWGN) with zero mean and variance $\sigma^{2}_{s,k}$ and $\sigma^{2}_{e,j}$. By employing PS to handle the received signal, the received signal at the ID circuit of $k$-th legitimate user  as
can be written as follows:
\begin{align}
	y^{ID}_{s,k}=\sqrt{\rho_{s,k}}(\sum_{l=1}^{K}\mathbf{h}_{s,k}^{H}\mathbf{q}_{l}+\mathbf{h}_{s,k}^{H}\mathbf{v}+n_{s.k})+n_{sp,k},
\end{align}
where $\rho_{s,k}\in(0,1]$ denotes the PS factor of the $k$-th legitimate user. In addition, $n_{sp,k}\sim \mathcal{CN}(0,\sigma_{sp,k}^{2})$ is the noise introduced by the ID circuit of the $k$-th legitimate user. The received SINR at the $k$-th legitimate receiver and the $j$-th eavesdroppers can be written as 
\begin{align}
\Gamma_{s,k}\!=\!\frac{\mathbf{h}_{s,k}^{H}\mathbf{q}_{k}\mathbf{q}_{k}^{H}\mathbf{h}_{s,k}}{\sum_{i\neq k}\mathbf{h}_{s,k}^{H}\mathbf{q}_{i}\mathbf{q}_{i}^{H}\mathbf{h}_{s,k}+\mathbf{h}_{s,k}^{H}\mathbf{V}\mathbf{h}_{s,k}+\sigma_{s,k}^{2}+\frac{\sigma_{sp,k}^{2}}{\rho_{s,k}}},
\end{align}
and
\begin{align}
\Gamma_{e,j}=\frac{\mathbf{h}_{e,j}^{H}\mathbf{q}_{k}\mathbf{q}_{k}^{H}\mathbf{h}_{e,j}}{\sum_{i\neq k}\mathbf{h}_{e,j}^{H}\mathbf{q}_{i}\mathbf{q}_{i}^{H}\mathbf{h}_{e,j}+\mathbf{h}_{e,j}^{H}\mathbf{V}\mathbf{h}_{e,j}+\sigma_{e,j}^{2}}.
\end{align}
On the other hand, the received signal for the EH circuit of the $k$-th legitimate user is given by
\begin{align}
y^{EH}_{s,k}=\sqrt{1-\rho_{s,k}}(\sum_{l=1}^{K}\mathbf{h}_{s,k}^{H}\mathbf{q_{l}}+\mathbf{h}_{s,k}^{H}\mathbf{v}+n_{s,k}).
\end{align}
The received RF power at the $k$-th legitimate user can be expressed as
\begin{align}
P_{k}\!=\!\zeta_{k}(1-\rho_{s,k})(\sum_{l=1}^{K}\mathbf{h}_{s,k}^{H}\mathbf{q}_{l}\mathbf{q}_{l}^{H}\mathbf{h}_{s,k}+\mathbf{h}_{s,k}^{H}\mathbf{V}\mathbf{h}_{s,k}+\sigma^{2}_{s,k}),
\end{align}
where $\zeta_{k}$ represents the EH efficiency of the $k$-th legitimate user. In this paper, we adopt a non-linear parametric EH model, which means the RF to DC conversion efficiency depends on the input power level. Then, the output DC power at the $k$-th legitimate user can be given by
\begin{align}
E_{k}=\frac{\Psi_{k}-M_{k}\Omega_{k}}{1-\Omega_{k}},
\end{align}
where $\Omega_{k}=\frac{1}{1+\textrm{exp}(a_{k}b_{k})}$,  $\Psi_{k}=\frac{M_{k}}{1+\textrm{exp}(-a_{k}(P_{k}-b_{k}))}$ and $M_{k}$ is a constant denoting the maximum output DC power at the $k$-th legitimate user. The parameter $a_{k}$ is the non-linear charging rate with the respect to the input power and $b_{k}$ is a minimum turn-on voltage of the EH circuit based parameter. 
\section{Problem Formulation}
As it is assumed that the eavesdroppers' CSI is unavailable at the transmitter, the best approach is to exploit AN to degrade intercepting capability of eavesdroppers. Our aim is to maximize the AN power while ensuring the worst-case SINRs and energy harvesting requirements are satisfied. This design problem to jointly optimize beamforming vectors and AN covariance matrix can be mathematically formulated as
	\begin{align}
	&\max_{\mathbf{q}_{k},\mathbf{V},\rho_{s,k}}  \textrm{tr}(\mathbf{V}) \nonumber\\
	s.t. &~  \Gamma_{s,k} \geq \gamma, E_{k}\geq \bar{E_{s}}, \forall k,\nonumber\\
	&~ \sum_{l=1}^{K}\|\mathbf{q}_{l}\|^{2} + \textrm{tr}(\mathbf{V}) \leq P_{\textrm{total}}, \nonumber\\
	&~ 0 < \rho_{s,k} \leq 1, \mathbf{V} \succeq \mathbf{0}. 
	\end{align}
Firstly, we tackle the information beamforming vector $\mathbf{q}_{k}$, by defining a new rank-one matrix $\mathbf{Q}_{k}=\mathbf{q}_{k}\mathbf{q}_{k}^{H}$. Then the original problem becomes
\begin{subequations}
	\begin{align}
	&\max_{\mathbf{Q}_{k},\mathbf{V},\rho_{s,k}}  \textrm{tr}(\mathbf{V}) \\
	s.t. &  \frac{\mathbf{h}_{s,k}^{H}\mathbf{Q}_{k}\mathbf{h}_{s,k}}{\sum_{i\neq k}\mathbf{h}_{s,k}^{H}\mathbf{Q}_{i}\mathbf{h}_{s,k}+\mathbf{h}_{s,k}^{H}\mathbf{V}\mathbf{h}_{s,k}+\sigma_{s,k}^{2}+\frac{\sigma_{sp,k}^{2}}{\rho_{s,k}}} \geq \gamma,\\
	 &~\zeta_{k}(1-\rho_{s,k})(\sum_{l=1}^{K}\mathbf{h}_{s,k}^{H}\mathbf{Q}_{l}\mathbf{h}_{s,k}+\mathbf{h}_{k}^{H}\mathbf{V}\mathbf{h}_{k}+\sigma^{2}_{s,k})\geq \omega_{k},\\
	&~ \textrm{tr}(\sum_{l=1}^{K}\mathbf{Q}_{l}+\mathbf{V}) \leq P_{\textrm{total}}, \mathbf{V} \succeq \mathbf{0}, \mathbf{Q}_{l} \succeq \mathbf{0}, \forall l, \label{eq:Sec_rate_max_power_constraints}\\
	&~ 0 < \rho_{s,k} \leq 1,  
	\textrm{rank}(\mathbf{Q}_{k})=1.\label{eq:Sec_rate_max_another_constraints}
	\end{align}
\end{subequations}
where $\omega_{k}$ represents the required received power under the non-linear EH model, which is given by
\begin{align}
\omega_{k}=b_{k}-\frac{\ln(\frac{M_{k}}{(\bar{E}_{s}+(M_{k}-\bar{E}_{s})\Omega_{k}}-1)}{a_{k}}.
\end{align}
By employing SDR, we relax problem (10) by dropping the rank constraint $\textrm{rank}(\mathbf{Q}_{k})=1$, the relaxed problem becomes:
\begin{align}
\max_{\mathbf{Q}_{k},\mathbf{V},\rho_{s,k}} &~ \textrm{tr}(\mathbf{V}) \nonumber\\
s.t. ~&  \textrm{(10b)-(10d)} \nonumber\\
&~ 0 < \rho_{s,k} \leq 1.
\end{align}

\begin{Proposition}\label{proposition:rank_proof}
	Provided that the problem (12) is feasible, the optimal solution will be always rank-one.
\end{Proposition}
\begin{IEEEproof}
Please refer to Appendix.
\end{IEEEproof}
\subsection{Robust Design with Ellipsoidal Channel Uncertainties}
In general, it is difficult to for the transmitter to obtain perfect CSI since channel estimation and quantization always have errors. As such, we assume that transmitter has imperfect CSI of the legitimate users. We model this imperfect CSI based on the deterministic models \cite{niu2017robust, ma2017cooperative} and the actual channel of the k th legitimate user can be written as follows: 
\begin{align}
\mathbf{h}_{s,k}=\mathbf{\hat{h}}_{s,k}+\mathbf{\hat{e}}_{s,k},
\end{align}
where $\mathbf{\hat{e}}_{s,k}$ represents the channel uncertainties. Based on deterministic model, the Euclidean norms of channel errors are bounded by a set of thresholds as
\begin{align}
||\mathbf{\hat{e}}_{s,k}||=||\mathbf{h}_{s,k}-\mathbf{\hat{h}}_{s,k}||\leq\epsilon_{s,k},
\end{align}
where $\epsilon_{s,k}\geq0$ denotes the upper bound of channel uncertainties. By incorporating these uncertainties, the original problem can be formulated as the following robust optimization problem:
\begin{subequations}
	\begin{align}
	&\max_{\mathbf{Q}_{k},\mathbf{V},\rho_{s,k}}  \textrm{tr}(\mathbf{V}) \\
	s.t. & (\mathbf{\hat{h}}_{s,k}+\mathbf{\hat{e}}_{s,k})^{H}\mathbf{Q}_{k}(\mathbf{\hat{h}}_{s,k}+\mathbf{\hat{e}}_{s,k})\geq\gamma[\sum_{i\neq k}(\mathbf{\hat{h}}_{s,k}+\mathbf{\hat{e}}_{s,k})^{H}\nonumber\\ &(\mathbf{Q}_{i}+\mathbf{V})(\mathbf{\hat{h}}_{s,k}+\mathbf{\hat{e}}_{s,k})+\sigma_{s,k}^{2}+\sigma_{sp,k}^{2}/\rho_{s,k}] ,\\
	&~\zeta(1-\rho_{s,k})\bigg[\sum_{l=1}^{K}(\mathbf{\hat{h}}_{s,k}+\mathbf{\hat{e}}_{s,k})^{H}(\mathbf{Q}_{l}+\mathbf{V})(\mathbf{\hat{h}}_{s,k}+\mathbf{\hat{e}}_{s,k})\nonumber\\
	&+\sigma^{2}_{k}\bigg]\geq \omega_{k}, \\
	&~ \textrm{tr}(\sum_{l=1}^{K}\mathbf{Q}_{l}+\mathbf{V}) \leq P_{\textrm{total}},\\  
	&~ 0 < \rho_{s,k} \leq 1, \mathbf{V} \succeq \mathbf{0}, \mathbf{Q}_{l} \succeq \mathbf{0},\forall l,\\
	&~\mathbf{h}_{s,k}=\mathbf{\hat{h}}_{s,k}+\mathbf{\hat{e}}_{s,k}, ||\mathbf{\hat{e}}_{s,k}||\leq\epsilon_{s,k}.	
	\end{align}
\end{subequations}
Then, we remove the channel errors and transform (15b) and (15c) into linear matrix inequality (LMI) forms. In order to convert (15b) into QMI form, we first define $\mathbf{W}_{k}=\frac{1}{\gamma}\mathbf{Q}_{k}-\sum_{i\neq k}\mathbf{Q}_{i}-\mathbf{V}$, then we employ the following lemma:

\emph{Lemma 1}: (\emph{Schur complement} \cite{boyd2004convex}): Let $\mathbf{X}$ be a complex hermitian matrix,
\begin{align}
\mathbf{X}=\mathbf{X}^{H}=\left[
\begin{matrix}
\mathbf{B}_{1} & \mathbf{B}_{2}\\
\mathbf{B}_{2}^{H} & \mathbf{B}_{3}\\
\end{matrix}\right].
\end{align}
Thus, $\mathbf{B}_{4}=\mathbf{B}_{3}-\mathbf{B}_{2}^{H}\mathbf{B}_{1}^{-1}\mathbf{B}_{2}$ is the Schur complement of $\mathbf{B}_{1}$ in $\mathbf{X}$ and the following statements holds: (1) $\mathbf{X}\succeq 0$, if and only if $\mathbf{B}_{1}\succeq 0$ and $\mathbf{B}_{4}\succeq 0$, (2) if $\mathbf{B}_{1}\succeq 0$ then $\mathbf{X}\succeq 0$ if and only if $\mathbf{B}_{4}\succeq 0$.\\

By utilizing Lemma 1, we can rewrite the constraint in (15b) as the following semidefinite constraint:
\begin{align}
\left[
\begin{matrix}
\rho_{s,k} & \sigma_{sp,k}\\
\sigma_{sp,k} &(\mathbf{\hat{h}}_{s,k}+\mathbf{\hat{e}}_{s,k})^{H}\mathbf{W}_{k}(\mathbf{\hat{h}}_{s,k}+\mathbf{\hat{e}}_{s,k})-\sigma_{s,k}^{2}\\
\end{matrix}\right]\succeq \mathbf{0}.
\end{align}
However, this semidefinite constraint in (17) is not convex due to the channel uncertainties $\hat{e}_{s,k}$. In order to make this constraint tractable, we use the following lemma to convert it into LMIs:

\emph{Lemma 2}: \cite{luo2004multivariate} If $\mathbf{D}\succeq 0$, then the following QMI
\begin{align}
\left[
\begin{matrix}
\mathbf{A}_{1} & \mathbf{A}_{2}+\mathbf{A}_{3}\mathbf{X}\\
(\mathbf{A}_{2}+\mathbf{A}_{3}\mathbf{X})^{H} &\mathbf{A}_{4}+\mathbf{X}^{H}\mathbf{A}_{5}+\mathbf{A}_{5}^{H}\mathbf{X}+\mathbf{X}^{H}\mathbf{A}_{6}\mathbf{X}\\
\end{matrix}\right]\succeq \mathbf{0},
\end{align}
\begin{align}
\forall \mathbf{X}: \mathbf{I}-\mathbf{X}^{H}\mathbf{D}\mathbf{X}\succeq \mathbf{0},
\end{align}
is equivalent to the following inequality. There exists $\lambda\geq 0$ such that
\begin{align}
\left[
\begin{matrix}
\mathbf{A}_{1} & \mathbf{A}_{2}  & \mathbf{A}_{3}\\
\mathbf{A}_{2}^{H} & \mathbf{A}_{4} & \mathbf{A}_{5}^{H}\\
\mathbf{A}_{3}^{H} & \mathbf{A}_{5} & \mathbf{A}_{6}
\end{matrix}\right]
-\lambda
\left[
\begin{matrix}
\mathbf{0} & \mathbf{0}  & \mathbf{0}\\
\mathbf{0} & \mathbf{I} & \mathbf{0}\\
\mathbf{0} & \mathbf{0} & -\mathbf{D}
\end{matrix}\right]\succeq \mathbf{0}.
\end{align}
 
 To proceed, we set $\mathbf{X}=\mathbf{\hat{e}}_{s,k}, \mathbf{D}=1/\epsilon_{s,k}^{2}\mathbf{I}, \mathbf{A}_{1}= \rho_{s,k}, \mathbf{A}_{2}=\sigma_{sp,k}, \mathbf{A}_{3}=\mathbf{0}_{1\times N_{T}}, \mathbf{A}_{4}=\mathbf{\hat{h}}_{s,k}^{H}\mathbf{W}_{k}\mathbf{\hat{h}}_{s,k}-\sigma_{s,k}, \mathbf{A}_{5}=\mathbf{W}_{k}\mathbf{\hat{h}}_{s,k}, \mathbf{A}_{6}=\mathbf{W}_{k}$. Now we rewrite the norm-bounded error vector $\mathbf{\hat{e}}_{s,k}$, i.e. $||\mathbf{\hat{e}}_{s,k}||\leq\epsilon_{s,k}$ as $\mathbf{I}-\mathbf{\hat{e}}_{s,k}^{H}\frac{\mathbf{I}}{\epsilon_{s,k}^{2}}\mathbf{\hat{e}}_{s,k}\succeq \mathbf{0}$, then adopting Lemma 2, the constraint (17) can be equivalently expressed as 
 \begin{align}
\left[
\begin{matrix}
\rho_{s,k} & \sigma_{sp,k} &\mathbf{0}_{1\times N_{T}}\\
\sigma_{sp,k} & \mathbf{\hat{h}}_{s,k}^{H}\mathbf{W}_{k}\mathbf{\hat{h}}_{s,k}-\sigma^{2}_{s,k}-\lambda_{k} & \mathbf{\hat{h}}_{s,k}^{H}\mathbf{W}_{k}\\
\mathbf{0}_{N_{T}\times 1} &\mathbf{W}_{k}\mathbf{\hat{h}}_{s,k} & \mathbf{W}_{k}+\frac{\lambda_{k}}{\epsilon_{s,k}^{2}}\mathbf{I}
\end{matrix}\right]\succeq \mathbf{0},
\end{align}
where $\lambda_{k}\geq 0$ is an auxiliary variable. Similarly, defining $\mathbf{M}_{k}=\sum_{l=1}^{K}\mathbf{Q}_{l}+\mathbf{V}$ and applying Lemma 1, the constraint in (15c) can be equivalently reformulated as
\begin{align}
\left[
\begin{matrix}
\zeta(1-\rho_{s,k}) & \sqrt{\omega_{k}}\\
\sqrt{\omega_{k}} & (\mathbf{\hat{h}}_{s,k}+\mathbf{\hat{e}}_{s,k})^{H}\mathbf{W}_{k}(\mathbf{\hat{h}}_{s,k}+\mathbf{\hat{e}}_{s,k})+\sigma_{s,k}^{2}
\end{matrix}\right]\succeq \mathbf{0}.
\end{align}
By setting $\mathbf{X}=\mathbf{\hat{e}}_{s,k}, \mathbf{D}=1/\epsilon_{s,k}^{2}\mathbf{I}, \mathbf{A}_{1}= \zeta(1-\rho_{s,k}), \mathbf{A}_{2}=\sqrt{\omega_{k}}, \mathbf{A}_{3}=\mathbf{0}_{1\times N_{T}}, \mathbf{A}_{4}=\mathbf{\hat{h}}_{s,k}^{H}\mathbf{M}_{k}\mathbf{\hat{h}}_{s,k}+\sigma_{s,k}, \mathbf{A}_{5}=\mathbf{M}_{k}\mathbf{\hat{h}}_{s,k}, \mathbf{A}_{6}=\mathbf{M}_{k}$ and employing Lemma 2, the constraint (22) can be recast as
\begin{align}
\left[
\begin{matrix}
\zeta(1-\rho_{s,k}) & \sqrt{\omega_{k}} &\mathbf{0}_{1\times N_{T}}\\
\sqrt{\omega_{k}}  & \mathbf{\hat{h}}_{s,k}^{H}\mathbf{M}_{k}\mathbf{\hat{h}}_{s,k}+\sigma^{2}_{s,k}-t_{k} & \mathbf{\hat{h}}_{s,k}^{H}\mathbf{M}_{k}\\
\mathbf{0}_{N_{T}\times 1} & \mathbf{M}_{k}\mathbf{\hat{h}}_{s,k} & \mathbf{M}_{k}+\frac{t_{s,k}}{\epsilon_{s,k}^{2}}\mathbf{I}
\end{matrix}\!\right]\!\succeq\! \mathbf{0},
\end{align}
where $t_{k}\geq 0$ is an auxiliary variable. Hence, the original problem (15) can be reformulated as
\begin{align}
&\max_{\mathbf{Q}_{k},\mathbf{V},\rho_{s,k}, \lambda_{k}, t_{k}}  \textrm{tr}(\mathbf{V}) \nonumber\\
s.t. &~ \textrm{(15d), (15e), (21), (23)},\nonumber\\
&~ \lambda_{k}\geq 0, t_{k}\geq 0. 
\end{align}
The problem (24) is convex and can be solved efficiently via CVX \cite{boyd2004convex}.
\subsection{Robust Design with Statistical Channel Uncertainties}
We next develop another robust design to handle statistical channel uncertainties which are modelled as Gaussian random variables with known statistical distributions. In particular, we assume that
\begin{align}
\mathbf{\hat{e}}_{s,k}\sim\mathcal{CN}(\mathbf{0},\mathbf{\Theta}_{s,k}),
\end{align}
where $\mathbf{\Theta_{s,k}}\succeq 0$ is the given covariance matrix for $\mathbf{\hat{e}}_{s,k}$. Then the robust problem can be written with statistical channel uncertainties as
\begin{subequations}
\begin{align}
&\max_{\mathbf{Q}_{k},\mathbf{V},\rho_{s,k}} ~ \textrm{tr}(\mathbf{V}) \\
s.t. &~ \textrm{Prob}(\Gamma_{s,k}\geq \gamma)\geq 1-p_{s,k}\\
&~\textrm{Prob}(E_{k}\geq \bar{E_{s}})\geq 1-q_{s,k},\\
&~ \mathbf{h}_{s,k}=\mathbf{\hat{h}}_{s,k}+\mathbf{\hat{e}}_{s,k}, \mathbf{\hat{e}}_{s,k}\sim\mathcal{CN}(\mathbf{0},\mathbf{\Theta}_{s,k}),\\
&~\textrm{(15d), (15e)}, 
\end{align}
\end{subequations}
 where $p_{s,k}\in (0,1]$ and $q_{s,k}\in (0,1]$ are the predefined outage probabilities for the SINR and EH requirements of the $k$-th legitimate user, respectively. These thresholds ensure that the $k$-th legitimate user is served with a satisfiable SINR and EH no less than $(1-p_{s,k})\times100\%$ and  $(1-q_{s,k})\times100\%$ of the time, respectively. The problem (26) is non-convex due to the probability constraints in (26a) and (26b) which do not have closed-form expressions. To circumvent this issue, we adopt the following lemma:
 
 \emph{Lemma 3}: (\emph{Berstein-type Inequality}): Let $\mathbf{e}\sim \mathcal{CN}(\mathbf{0},\mathbf{I}_{n})$, $\mathbf{Q}\in \mathbb{H}^{n}$, $\mathbf{r}\in \mathbb{C}^{n}$ and $s\in\mathbb{R}$, for any $0<p\leq 1$, we have
 \begin{align}
 \textrm{Prob}\{\mathbf{e}^{H}\mathbf{Q}\mathbf{e}+2Re(\mathbf{e}^{H}\mathbf{r})+s\geq 0\} \geq 1-p,
 \end{align}
is equivalent to the following set of convex constraints:
\begin{align}
\textrm{tr}(\mathbf{Q})-\sqrt{-2\textrm{ln}(p)}t_{1}+\textrm{ln}(p)t_{2}+s\geq 0,\nonumber\\
\bigg|\bigg|\left[
\begin{matrix}
\textrm{vec}(\mathbf{Q})  \\
\sqrt{2}\mathbf{r}
\end{matrix}\right]\bigg|\bigg|\leq t_{1},\nonumber\\
t_{2}\mathbf{I}+\mathbf{Q}\succeq \mathbf{0}, t_{2}\geq 0，
\end{align}
where $t_{1}$ and $t_{2}$ are slack variables. We first define the following equations
\begin{align}
\mathbf{B}_{k}=\mathbf{\Theta}_{s,k}^{\frac{1}{2}}\mathbf{W}_{k}\mathbf{\Theta}_{s,k}^{\frac{1}{2}}, \mathbf{r}_{k}=\mathbf{\Theta}_{s,k}^{\frac{1}{2}}\mathbf{W}_{k}\mathbf{\hat{h}}_{s,k},\\
s_{k}=\mathbf{\hat{h}}_{s,k}^{H}\mathbf{W}_{k}\mathbf{\hat{h}}_{s,k}-\sigma^{2}_{s,k}-\sigma^{2}_{sp,k}/\rho,\\
\mathbf{E}_{k}=\mathbf{\Theta}_{s,k}^{\frac{1}{2}}\mathbf{M}_{k}\mathbf{\Theta}_{s,k}^{\frac{1}{2}}, \mathbf{g}_{k}=\mathbf{\Theta}_{s,k}^{\frac{1}{2}}\mathbf{M}_{k}\mathbf{\hat{h}}_{s,k},\\
l_{k}=\mathbf{\hat{h}}_{s,k}^{H}\mathbf{M}_{k}\mathbf{\hat{h}}_{s,k}+\sigma^{2}_{s,k}-\frac{\omega_{k}}{\zeta(1-\rho_{k})}.
\end{align}
Then, by applying Lemma 3 and Lemma 1, the constraint in (26b) is equivalent to
\begin{align}
\left[
\begin{matrix}
f_{s,k}+\mathbf{\hat{h}}_{s,k}^{H}\mathbf{W}_{k}\mathbf{\hat{h}}_{s,k}-\sigma^{2}_{s,k}&\sigma_{sp,k}  \\
\sigma_{sp,k} & \rho_{s,k}
\end{matrix}\right]\succeq \mathbf{0},\nonumber\\
\bigg|\bigg|\left[
\begin{matrix}
\textrm{vec}(\mathbf{B}_{k})  \\
\sqrt{2}\mathbf{r}_{k}
\end{matrix}\right]\bigg|\bigg|\leq x_{s,k},\nonumber\\
y_{s,k}\mathbf{I}+\mathbf{B}_{k}\succeq \mathbf{0}, y_{s,k}\geq 0,
\end{align}
where $f_{s,k}=\textrm{tr}(\mathbf{B}_{k})-\sqrt{-2\textrm{ln}(p_{s,k})}x_{s,k}+\textrm{ln}(p_{s,k})y_{s,k}$. Similarity, the constraint in (26c) can be equivalently written as
\begin{align}
\left[
\begin{matrix}
g_{s,k}+\mathbf{\hat{h}}_{s,k}^{H}\mathbf{M}_{k}\mathbf{\hat{h}}_{s,k}+\sigma^{2}_{s,k}&\sqrt{\omega_{k}}  \\
\sqrt{\omega_{k}} & \zeta_{k}(1-\rho_{s,k})
\end{matrix}\right]\succeq \mathbf{0},\nonumber\\
\bigg|\bigg|\left[
\begin{matrix}
\textrm{vec}(\mathbf{E}_{k})  \\
\sqrt{2}\mathbf{g}_{k}
\end{matrix}\right]\bigg|\bigg|\leq m_{s,k},\nonumber\\
n_{s,k}\mathbf{I}+\mathbf{E}_{k}\succeq \mathbf{0}, n_{s,k}\geq 0,
\end{align}
where $g_{s,k}=\textrm{tr}(\mathbf{E}_{k})-\sqrt{-2\textrm{ln}(q_{s,k})}m_{s,k}+\textrm{ln}(q_{s,k})n_{s,k}$. The statistical channel uncertainty based robust optimization problem in (26) is reformulated as 
\begin{align}
&\max_{\mathbf{Q}_{k},\mathbf{V},\rho_{s,k}} ~ \textrm{tr}(\mathbf{V})\nonumber \\
s.t. &~\textrm{(15d), (15e), (33), (34)}.
\end{align}
The formulation in (35) is a standard SDP and can be solved efficiently by CVX \cite{boyd2004convex}.
\section{Simulation Results}
In this section, we present numerical results to validate the performance of our proposed robust schemes. Here, we consider a wireless communication system with one transmitter and three legitimate users, where the transmitter is equipped with $N_{T}=4$ antennas and all the legitimate users are equipped with single antenna. It is assumed that all the channel models include both large scale and small scale fading. The simplified large scale fading model is given by $D=(\frac{d}{d_{0}})^{-\alpha}$, where $d$ is the distance between the transmitter and the receiver, we set $d_{0}=10 m$ as the reference distance, and $\alpha=2$ as the path loss exponent. We assume that $d_{1}$=10 m, $d_{2}$=14 m and $d_{3}$ =18 m as the distance between the transmitter and the 1st, 2nd and 3rd legitimate user, respectively. We model all the channel coefficients as Rician fading, which can be expressed as 
\begin{align}
\mathbf{\hat{h}}_{s,k}=\sqrt{\frac{K_{R}}{1+K_{R}}}\mathbf{\hat{h}}_{s,k}^{LOS}+\sqrt{\frac{1}{1+K_{R}}}\mathbf{\hat{h}}_{s,k}^{NLOS},
\end{align}
where $\mathbf{\hat{h}}_{s,k}^{LOS}$ indicates the line of sight (LOS) deterministic component with $||\mathbf{\hat{h}}_{s,k}^{LOS}||^{2}_{2}=D$, $\mathbf{\hat{h}}_{s,k}^{NLOS}\sim \mathcal{CN}(\mathbf{0},D\mathbf{I})$ represents the Rayleigh fading component. $K_{R}$ is the Rician factor which is set to be 3. In addition, for the LOS component, we employ the far-field uniform linear antenna array to model the channels \cite{karipidis2007far}.  For the non-linear EH model, we set $M$ to be 10 dBm which corresponds to the maximum output DC power. Besides, we adopt $a_{k}=150~ \forall k$ and $b_{k}=0.024~ \forall k$ as provided in \cite{boshkovska2016power, guo2012improved}. The available total transmit power $P_{total}$ is assumed to be 30 dBm. The EH efficiency $\zeta_{k}$ is set to be 1. Note that the CSI of the eavesdroppers is unavailable in both robust designs. To determine the worst-case SINR for eavesdroppers, we randomly generate eavesdroppers' channels in the same way as the legitimate users' channels, by setting all the distance between the transmitter and eavesdroppers to $8$ m and the number of eavesdroppers $J=2$. For the statistical channel uncertainty based robust design, the covariance matrices are set to be $\Omega_{k}=\epsilon_{s,k}^{2}\mathbf{I}, ~\forall k$. In addition, the SINR and EH outage probability requirements in this design are set to $p_{s,k}=q_{s,k}=0.1$. All the parameters described above are used in simulations unless specified.
\begin{center}
	\begin{figure}[ht!]
		\includegraphics[width=\linewidth]{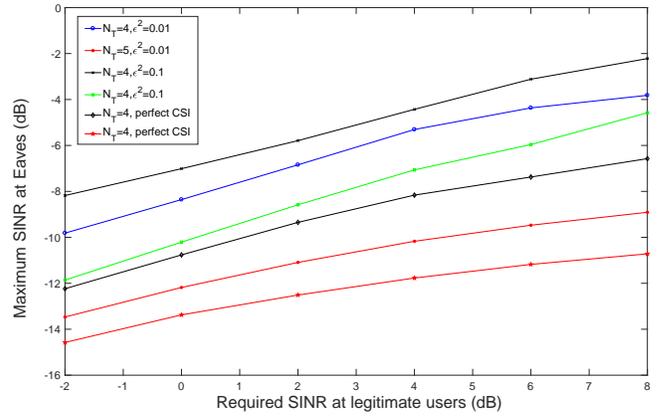}
		\caption{Maximum SINR among eavesdroppers versus the SINR requirement at legitimate users with different antennas and error bounds for the robust design with ellipsoidal channel uncertainties.}
	\end{figure}
\end{center}
\begin{center}
	\begin{figure}[ht!]
		\includegraphics[width=\linewidth]{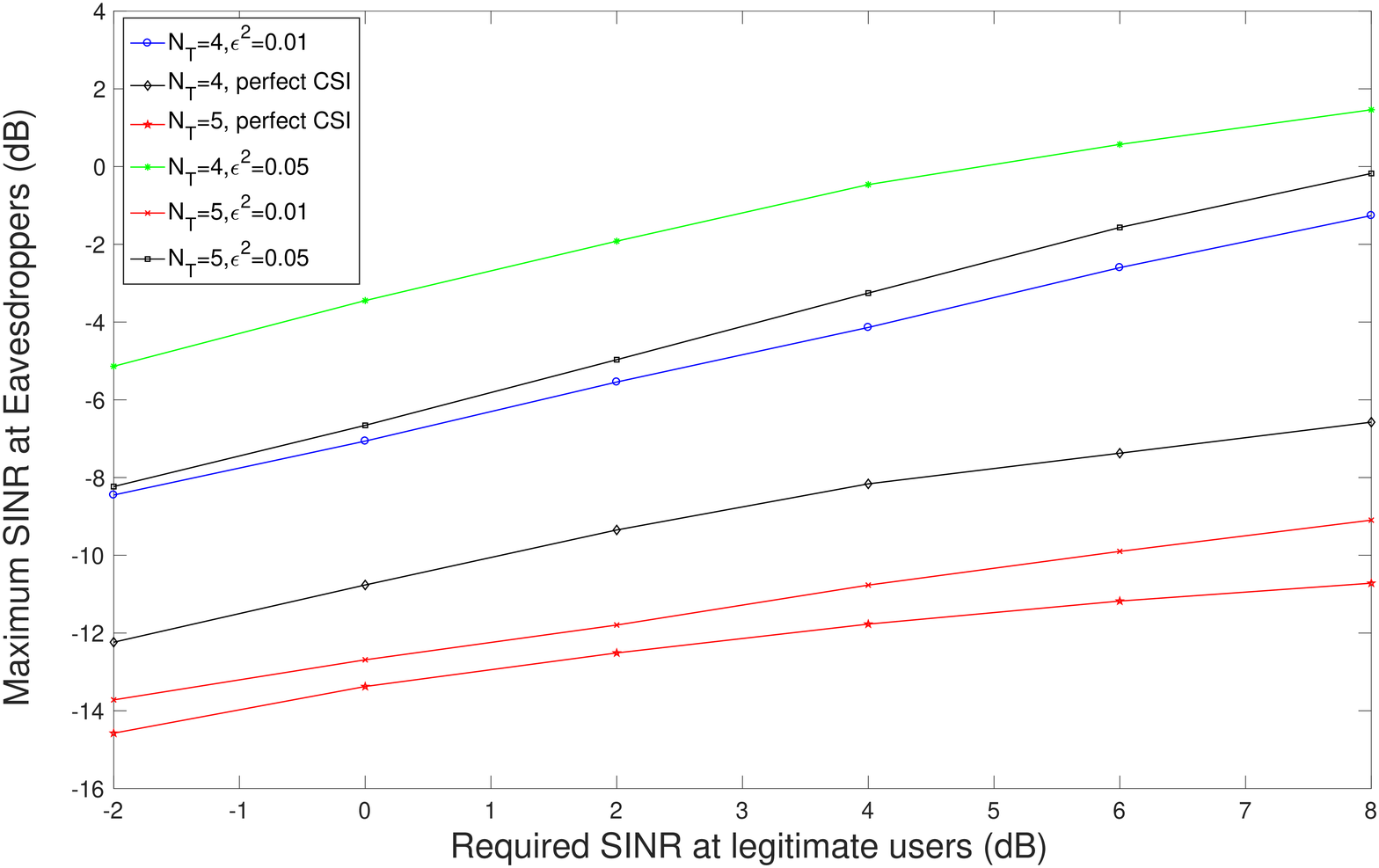}
		\caption{Maximum SINR among eavesdroppers versus the SINR requirement at legitimate users with different antennas and error bounds for the robust design with statistical channel uncertainties.}
	\end{figure}
\end{center}

Fig. 2 compares achieved maximum SINR at the eavesdroppers between the perfect CSI design and robust design with the ellipsoidal channel uncertainties for different target SINRs and EH $(\bar{E_s} = 8~\textrm{dBm})$ requirements. As seen in Fig. 2, the largest SINR at the eavesdroppers increases with the SINR targets and the error bounds. The design with perfect CSI achieves the least SINR at the eavesdroppers with the same number of transmit antennas. Besides, it is obvious that the achievable SINR at the eavesdroppers can be reduced with large number of transmit antennas due to more degrees of freedom in the beamforming design.

Fig. 3 compares achieved maximum SINR at the eavesdroppers between the perfect CSI design and the robust design with statistical channel uncertainties for different target SINRs and EH $(\bar{E_s} = 8~\textrm{dBm})$ requirements. As seen in Fig. 3, the largest SINR at the eavesdroppers increases with the SINR targets and the error bounds. The design with perfect CSI achieves the least SINR at the eavesdroppers with the same number of transmit antennas. Similar to Fig. 2, it is obvious that the achievable SINR at the eavesdroppers can be reduced by using a large number of transmit antennas.

\begin{center}
	\begin{figure}[ht!]
		\includegraphics[width=\linewidth]{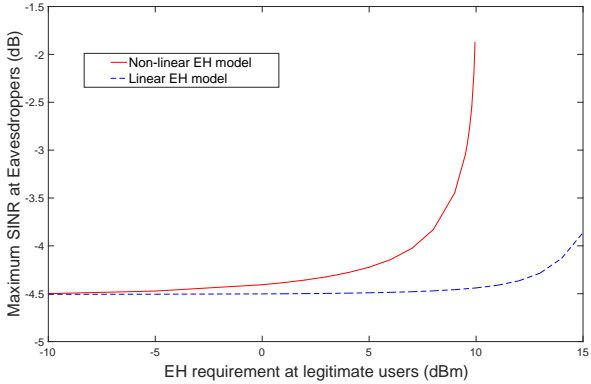}
		\caption{Non-linear EH model vs linear EH model for robust design with ellipsoidal channel uncertainties.}
	\end{figure}
\end{center}
\begin{center}
	\begin{figure}[ht!]
		\includegraphics[width=\linewidth]{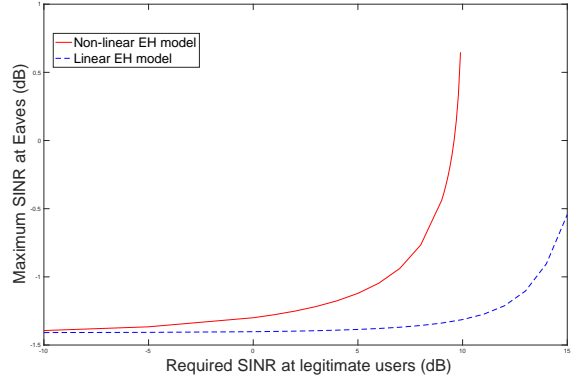}
		\caption{Non-linear EH model vs linear EH model for robust design with statistical channel uncertainties.}
	\end{figure}
\end{center}

Fig. 4 compares non-linear EH model and linear EH model for robust design with ellipsoidal channel uncertainties. The EH requirement is set to be $8$ dBm with the error-bound $\epsilon^{2}=0.01$. The largest SINR at the eavesdroppers increases as EH requirement increases for both EH models. However, there exists a saturation point on the EH requirement $\bar{E_{s}}$ in the non-linear EH model, as seen in Fig. 4, due to the non-linear characteristics of practical RF-EH conversion circuit. Moreover, for the non-linear EH model, the EH requirements can be guaranteed only with $\bar{E}_{s} < M$ due to the maximum output DC power limitation of the practical RF-EH conversion circuit. On the other hand, adopting the linear EH model may lead to false output DC power. Hence, employing the non-linear EH model can reflect the characteristics of the practical RF-EH conversion circuit. Similar performance is shown in Fig. 5 for the statistical channel uncertainty based robust design with $\bar{E_{s}}=8$ dBm, and $\epsilon^{2}=0.01$.
\section{Conclusions}
This work studied different robust beamforming designs in an AN-aided MISO SWIPT system in the presence of multiple purely passive eavesdroppers. By considering practical non-linear EH models, we have developed two robust designs based on different types of channel uncertainties: ellipsoidal and statistically based channel uncertainties. In the robust design with ellipsoidal uncertainties, we reformulated the original problem as a SDP through a linear matrix inequality representation. By exploiting both SDR and Berstein type inequality, we showed that robust design with statistical channel uncertainties can be recast as another SDP. Simulation results were provided to demonstrate the performance of both robust designs. Besides, the performance of the linear and non-linear EH model have been compared and by adopting the non-linear EH model can avoid false output power.

\begin{appendix}
	\subsection*{Proof of Proposition \ref{proposition:rank_proof}}\label{proof_of_proposition}
	First, we consider the Lagrangian function of problem (12):
	\begin{align}
		&\mathcal{L}(\mathbf{Q}_{k},\mathbf{V}, \mathbf{Z}_{k}, \mathbf{Y}, \alpha,\beta,\gamma,\mu)\!=\!-\textrm{tr}(\mathbf{V}-\textrm{tr}(\mathbf{Z}_{k}\mathbf{Q}_{k})-\textrm{tr}(\mathbf{Y}\mathbf{V})\nonumber\\	&\!-\!\lambda_{k}\{\textrm{tr}[\mathbf{h}_{s,k}\mathbf{h}_{s,k}^{H}(\frac{\mathbf{Q}_{k}}{\gamma}-\sum_{i\neq k}\mathbf{Q}_{i}- \mathbf{V})- \sigma_{s,k}^{2}- \sigma_{sp,k}^{2}/\rho_{s,k}]\} \nonumber\\
	 &\!-\!\mu_{k}\{\textrm{tr}[\mathbf{h}_{s,k}\mathbf{h}_{s,k}^{H}(\sum_{l=1}^{K}\mathbf{Q}_{l}+\mathbf{V})+\sigma_{s,k}^{2}]-\omega_{k}/[\zeta(1-\rho_{s,k})]\}\nonumber\\
	 &+\alpha[\textrm{tr}(\sum_{l=1}^{K}\mathbf{Q}_{l}+\mathbf{V})-P_{total}],
	\end{align}
	where $\mathbf{Z}_{k}\in \mathbb{H}^{N_{T}}_{+}$, $\mathbf{Y} \in \mathbb{H}^{N_{T}}_{+}$, $\lambda_{k}\in \mathbb{R}_{+}$, $\mu_{k}\in\mathbb{R}_{+}$, $\alpha\in\mathbb{R}_{+}$ are the Lagrangian multipliers associated with problem (12). Then we derive the corresponding  Karush-Kuhn-Tucker (KKT) conditions \cite{boyd2004convex}:
	\begin{align}
		&\frac{\partial\mathcal{L}}{\partial\mathbf{Q}_{k}}=-\mathbf{Z}_{k}-(\lambda_{k}/\gamma+\mu_{k})\mathbf{h}_{s,k}\mathbf{h}_{s,k}^{H}+\alpha\mathbf{I}=\mathbf{0},\\
		&\frac{\partial\mathcal{L}}{\partial\mathbf{V}}=-\mathbf{I}-\mathbf{Y}+(\lambda_{k}-\mu_{k})\mathbf{h}_{s,k}\mathbf{h}_{s,k}^{H}+\alpha\mathbf{I}=\mathbf{0},\\
		&\mathbf{Z}_{k}\mathbf{Q}_{k}=\mathbf{0}, \mathbf{Z}_{K}\succeq \mathbf{0}, \mathbf{Y}\mathbf{V}=\mathbf{0}, \mathbf{Y}\succeq \mathbf{0}.
	\end{align}
	The following equality holds:
	\begin{align}
		&\textrm{(38)-(39)}= -\mathbf{Z}_{k}+\mathbf{I}+\mathbf{Y}-\lambda_{k}(1+\frac{1}{\gamma})\mathbf{h}_{s,k}\mathbf{h}_{s,k}^{H}=\mathbf{0},
		\end{align}
	\begin{align}
		\Rightarrow \mathbf{Z}_{k}=\mathbf{I}+\mathbf{Y}-\lambda_{k}(1+\frac{1}{\gamma})\mathbf{h}_{s,k}\mathbf{h}_{s,k}^{H},
	\end{align}
	\begin{align}
	\Rightarrow
		\{\mathbf{I}+\mathbf{Y}-\lambda_{k}(1+\frac{1}{\gamma})\mathbf{h}_{s,k}\mathbf{h}_{s,k}^{H}\}\mathbf{Q}_{k}=\mathbf{0}.
	\end{align}
	Hence, the following rank relation holds:
	\begin{align}
		&\textrm{rank}(\mathbf{Q}_{k})=\textrm{rank}\bigg\{\frac{\lambda_{k}(1+\frac{1}{\gamma})\mathbf{h}_{s,k}\mathbf{h}_{s,k}^{H}}{\mathbf{I}+\mathbf{Y}}\mathbf{Q}_{k}\bigg\}\nonumber\\ &\leq\textrm{rank}[\lambda_{k}(1+\frac{1}{\gamma})\mathbf{h}_{s,k}\mathbf{h}_{s,k}^{H}]\leq 1,
	\end{align}
	which completes the proof of proposition 1.		
\end{appendix}


%\section*{Acknowledgment}


%\bibliographystyle{IEEEtran}
\bibliographystyle{IEEEtran}
\bibliography{referenceIEEE}

\end{document}